\documentclass[aps,twocolumn,superscriptaddress]{revtex4}
\usepackage{graphicx}

\newcommand{\ket}[1] {\left| #1 \right\rangle}

\begin{document}

\title{Speed limits for quantum gates in multi-qubit systems}

\author{S. Ashhab}
\affiliation{Advanced Science Institute, RIKEN, Wako-shi, Saitama
351-0198, Japan}
\affiliation{Physics Department, The University of Michigan, Ann
Arbor, Michigan 48109-1040, USA}

\author{P. C. de Groot}
\affiliation{Kavli Institute of Nanoscience, Delft University of
Technology, P.O. Box 5046, 2600 GA Delft, The Netherlands}
\affiliation{Present address: Max-Planck-Institut f\"ur
Quantenoptik, 85748 Garching, Germany}
%

\author{Franco Nori}
\affiliation{Advanced Science Institute, RIKEN, Wako-shi, Saitama
351-0198, Japan}
\affiliation{Physics Department, The University of Michigan, Ann
Arbor, Michigan 48109-1040, USA}

\date{\today}


\begin{abstract}
We use analytical and numerical calculations to obtain speed
limits for various unitary quantum operations in multi-qubit
systems under typical experimental conditions. The operations that
we consider include single-, two- and three-qubit gates, as well
as quantum state transfer in a chain of qubits. We find in
particular that simple methods for implementing two-qubit gates
generally provide the fastest possible implementations of these
gates. We also find that the three-qubit Toffoli gate time varies
greatly depending on the type of interactions and the system's
geometry, taking only slightly longer than a two-qubit
controlled-NOT (CNOT) gate for a triangle geometry. The speed
limit for quantum state transfer across a qubit chain is set by
the maximum spin wave speed in the chain.
\end{abstract}


\maketitle

\section{Introduction}
\label{Sec:Introduction}

There are a number of candidate physical systems for the
implementation of qubits in a future quantum computer
\cite{QubitReviews}. Single-, two- and three-qubit gates have been
implemented in some of these systems in the past few years.
Various other quantum-information-related tasks based on
collective manipulation of qubits have also been demonstrated on
larger systems \cite{QubitReviews}.

As qubit systems advance towards large-scale demonstrations and
practical applications, it becomes increasingly important to
optimize the time required to implement the different operations,
such that the maximum number of operations is achieved within the
coherence time of the system. This goal is the main motivation of
this work.

The question of time-optimal control has already been discussed in
a number of studies in the literature. For example,
Refs.~\cite{Barenco,Yuan} considered optimized constructions of
general quantum gates using sequences of basic gates. References
\cite{MandelstamTamm} discussed the relationship between speed
limits on quantum gates and the different energy scales in a
physical system, while Refs.~\cite{GeometricArguments} explored
the analogy between the problem of finding the minimum times for
quantum gates and the problem of finding geodesics in curved
spaces. These ideas were applied in Ref.~\cite{Vidal} in order to
find a general recipe for calculating speed limits for two-qubit
gates. The time-optimal implementation of the quantum Fourier
transform in multi-qubit systems was analyzed in
Ref.~\cite{SchulteHerbrueggen}, and the time-optimal
implementation of the CNOT gate on indirectly coupled qubits was
studied in Ref.~\cite{Carlini}. References
\cite{Spoerl,Stojanovic} performed numerical calculations in order
to determine the minimum time required for quantum gates in
specific experimental setups based on superconducting qubits. The
speed limit on quantum state transfer in long spin chains has also
been studied recently \cite{Osborne,Caneva}.

Here we consider a number of important operations in a variety of
possible setups, with varying degrees of single-qubit control and
inter-qubit coupling mechanisms. We use analytical arguments and
numerical calculations based on optimal control theory in order to
give speed limits for these operations.

The paper is organized as follows: In Sec.~\ref{Sec:Hamiltonians}
we introduce the different possible setups, with varying forms of
single-qubit controls and interactions. We then discuss the speed
limits of single-qubit gates (Sec.~\ref{Sec:SingleQubitGates}),
two-qubit gates (Sec.~\ref{Sec:TwoQubitGates}) and three-qubit
gates (Sec.~\ref{Sec:ThreeQubitGates}). In
Sec.~\ref{Sec:StateTransfer} we discuss the problem of quantum
state transfer. We conclude with a brief summary of the results in
Sec.~\ref{Sec:Conclusion}.

\section{Different types of qubits and their coupling}
\label{Sec:Hamiltonians}

Over the years, various physical systems and designs have been
proposed and demonstrated as implementations of qubits. This
variety means that the degree of control in qubit manipulation and
the physical mechanisms for coupling between qubits vary from one
system to another. In some cases, the qubit is formed by the
lowest two energy levels of a multi-level quantum system, adding
complications to the control requirements of the qubit. Here we
shall focus on ``good'' qubits, where a description with only two
quantum states provides a good approximation of the physical
system.

With the assumption of two-state qubits, the single-qubit
Hamiltonian can be expressed in terms of the two-dimensional Pauli
matrices $\hat{\sigma}_{\alpha}$ (with $\alpha=x$, $y$ or $z$):
\begin{equation}
\hat{H} = -\frac{\Delta(t)}{2} \hat{\sigma}_x -
\frac{\epsilon(t)}{2} \hat{\sigma}_z,
\label{Eq:SingleQubitHamiltonian}
\end{equation}
where the time dependence in Eq.~(\ref{Eq:SingleQubitHamiltonian})
suggests that both $\Delta$ and $\epsilon$ are tunable. Some
experimental setups (e.g.~early experiments on superconducting
qubits) have only one tunable parameter, typically expressed as
$\epsilon$ in Eq.~(\ref{Eq:SingleQubitHamiltonian}). We shall
consider both cases below.

Driving signals used for the manipulation of the qubits can be
applied through the tunable parameters in the Hamiltonian. We
shall assume that any arbitrary driving signal can be applied to
the system. In other words, we look for the fastest implementation
of quantum operations in the space of all possible control
signals.

The coupling Hamiltonian between two qubits is typically of Ising
or Heisenberg form. The former is described by the Hamiltonian
\begin{equation}
\hat{H}_{\rm I} = J \hat{\sigma}_z^{(i)} \otimes
\hat{\sigma}_z^{(j)},
\label{Eq:IsingCouplingHamiltonian}
\end{equation}
while the latter is described by the Hamiltonian
\begin{equation}
\hat{H}_{\rm H} = J \left( \hat{\sigma}_x^{(i)} \otimes
\hat{\sigma}_x^{(j)} + \hat{\sigma}_y^{(i)} \otimes
\hat{\sigma}_y^{(j)} + \hat{\sigma}_z^{(i)} \otimes
\hat{\sigma}_z^{(j)} \right).
\label{Eq:HeisenbergCouplingHamiltonian}
\end{equation}
where $J$ is the coupling strength and the superscripts $i$ and
$j$ denote the two coupled qubits. There are situations where the
coupling strength is tunable, e.g.~using additional coupler
elements in the system. However, since we are interested in the
speed limits for performing multi-qubit gates, we shall assume
that one would want to set $J$ at its maximum achievable value and
therefore treat $J$ as a fixed parameter in the calculations
below. It is worth mentioning here that it is possible in
principle to have fixed values of $\Delta$ and $\epsilon$ and
still be able to obtain the desired gates via the modulation of
$J$, an approach sometimes referred to as parametric coupling
\cite{Bertet}.

The parameters of the single-qubit Hamiltonian are typically much
larger than the inter-qubit coupling strength,
i.e.~$J\ll\Delta,\epsilon$. This separation in energy scales
simplifies the process of identifying the central elements in the
speed limits found in our calculations, and it makes the results
easily applicable to different setups.

\section{Single-qubit gates}
\label{Sec:SingleQubitGates}

Since the parameters of the single-qubit Hamiltonian are typically
much larger than the inter-qubit coupling strength, one can ignore
inter-qubit interactions when performing single-qubit gates.
Furthermore, performing single-qubit gates typically requires a
negligibly short duration compared to the duration required for
performing a two- or multi-qubit gate, such that the time required
for performing single-qubit gates is usually ignored for purposes
of evaluating the computational cost of a given multi-qubit task.

Ignoring interactions, and thus reducing the problem of finding
optimal pulses and speed limits for performing a given
single-qubit gate to a single-qubit problem, the task at hand
becomes straightforward. A common situation is that in which one
of the two parameters, say $\epsilon$, is tunable over a much
larger range than the other one, while the other parameter is
either fixed or tunable over a much smaller range. A rotation by
an angle $\beta$ about an axis that makes an angle $\theta$ with
the $z$ axis and an angle $\phi$ with the $xz$ plane can be
implemented as follows: rotate the state by an angle $-\phi$ about
the $z$ axis, set the Hamiltonian to
$\Delta(\hat{\sigma}_x+\hat{\sigma}_z\cot\theta)/2$ and let it act
for a duration $\beta\sin\theta/\Delta$, and finally rotate the
state by an angle $\phi$ about the $z$ axis. The first and last
steps are fast operations that are implemented by setting
$\epsilon$ to a value that is much larger than $\Delta$. As a
result, the duration of the second step is the limiting factor for
the minimum time required for implementing the desired rotation.
One can therefore say that the speed limit is set by the smaller
of the two qubit parameters (or more accurately the smaller of the
largest achievable values of the two parameters), which in the
above example is $\Delta$.

\section{Two-qubit gates}
\label{Sec:TwoQubitGates}

We start the discussion of finding the speed limits for two-qubit
gates by mentioning two approaches that might seem promising at
first sight, but to our knowledge are not always applicable to the
problem at hand. First, there are expressions for the speed limits
of quantum operations based on the energy and the spread in energy
of the quantum state \cite{MandelstamTamm}. The reason why these
arguments do not apply straightforwardly here can be seen by
considering two qubits with a coupling strength that is much
smaller than the inter-qubit detuning. The energy scales of the
combined system are then, to a very good approximation, set by the
individual qubit energies and their detuning from each other. The
coupling strength only slightly modifies the energy eigenstates
and eigenvalues. The coupling strength must, however, be the
limiting factor for performing two-qubit gates. The larger energy
scales can be used to set a lower bound on the required gate time
(i.e.~an upper bound on the speed). However, the coupling strength
would give a much higher lower bound.

The other approach is that in which two-qubit gates are visualized
using geometric representations, and the process of performing a
two-qubit gate is seen as the motion of the system's propagator
through the space of all two-qubit quantum operations
\cite{GeometricArguments}. This approach does indeed provide an
intuitive view of the problem and can be very useful in
calculations. Furthermore, for the case of fully tunable
single-qubit parameters, i.e.~the case where one can assume to
have (i) a fixed interaction Hamiltonian with no single-qubit
terms and (ii) the ability to perform fast single-qubit gates,
Ref.~\cite{Vidal} gives a recipe for determining the speed limit
of any two-qubit gate with any interaction Hamiltonian. However,
it is not obvious that the speed limits provided by this approach
apply to the case of fixed $\Delta$, which is relevant to a good
number of realistic experiments.

Four representative gates that are commonly studied in the
literature are: iSWAP, controlled-Z (CZ), CNOT and $\sqrt{\rm
SWAP}$. There are a number of simple methods that can be obtained
using intuition for the implementation of these gates assuming a
given (fixed) coupling strength $J$ and tunable single-qubit
parameters. The basic techniques, which can also be shown to be
time-optimal \cite{Vidal}, are summarized in the following
\cite{Schuch}:

\textbf{iSWAP Gate}: The standard approach to implementing the
iSWAP gate with Ising interactions is to put the two qubits in
resonance with each other, i.e.~setting
$\Delta_1=\Delta_2=\Delta$, with $\epsilon_1=\epsilon_2=0$ and
$J\ll\Delta$:
\begin{eqnarray}
\hat{H} & = & - \frac{\Delta_1}{2} \hat{\sigma}_x^{(1)} -
\frac{\Delta_2}{2} \hat{\sigma}_x^{(2)} + J \hat{\sigma}_z^{(1)}
\otimes \hat{\sigma}_z^{(2)} \nonumber
\\
& \approx & - \frac{\Delta_1}{2} \hat{\sigma}_x^{(1)} -
\frac{\Delta_2}{2} \hat{\sigma}_x^{(2)} + J \left(
\hat{\sigma}_+^{(1)} \otimes \hat{\sigma}_-^{(2)} +
\hat{\sigma}_-^{(1)} \otimes \hat{\sigma}_+^{(2)} \right)
\nonumber \\
\label{Eq:iSWAPHamiltonian}
\end{eqnarray}
where the operators $\hat{\sigma}_{\pm}$ are raising and lowering
operators that excite or de-excite the individual qubits between
their single-qubit energy eigenstates, which in this case are the
eigenstates of $\hat{\sigma}_x$. When allowed to act for a
duration $t=\pi/(2J)$, the Hamiltonian in
Eq.~(\ref{Eq:iSWAPHamiltonian}) effects an iSWAP gate, in addition
to two single-qubit rotations.

\textbf{Controlled-$\pi$-phase gate}: With Ising interactions, the
controlled-$\pi$-phase (or CZ) gate can be performed by setting
$\Delta_1=\Delta_2=0$:
\begin{equation}
\hat{H} = - \frac{\epsilon_1}{2} \hat{\sigma}_z^{(1)} -
\frac{\epsilon_2}{2} \hat{\sigma}_z^{(2)} + J \hat{\sigma}_z^{(1)}
\otimes \hat{\sigma}_z^{(2)},
\label{Eq:CPhaseHamiltonian}
\end{equation}
with no conditions on $\epsilon_1$ and $\epsilon_2$. When allowed
to act for a duration $t=\pi/(4J)$, the Hamiltonian in
Eq.~(\ref{Eq:CPhaseHamiltonian}) effects a CZ gate, in addition to
two single-qubit rotations and an (unimportant) overall phase
factor.

\textbf{CNOT gate}: The CNOT gate can be obtained by combining the
CZ gate with two single-qubit gates. These are $\pi/2$ pulses
applied to the target qubit before and after the CZ gate. As a
result, the amount of time required to obtain the CNOT gate using
this approach is approximately equal to the amount of time
required for the CZ gate, i.e.~$\pi/(4J)$ with Ising interactions.

\textbf{$\sqrt{\rm \bf SWAP}$ gate}: The $\sqrt{\rm SWAP}$ gate is
more naturally obtained with Heisenberg interactions. In this
case, one sets $\Delta_1=\Delta_2$ and $\epsilon_1=\epsilon_2$,
and after a time $t=\pi/(8J)$ one obtains the $\sqrt{\rm SWAP}$
gate.

\begin{figure}[h]
\includegraphics[width=8.0cm]{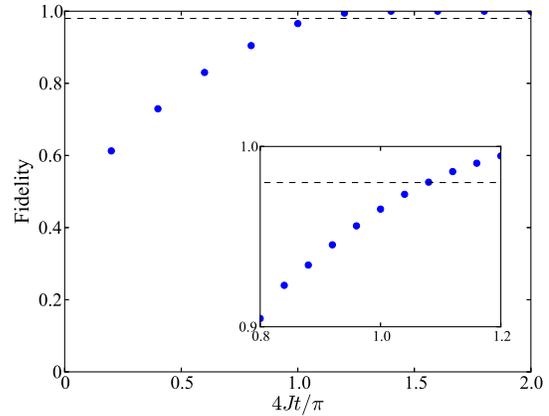}
\caption{(Color online) The fidelity of the CNOT gate for the
pulse obtained using optimal control theory as a function of the
allowed time $t$ (plotted in the combination $4Jt/\pi$) in the
case of Ising interactions and fixed values of $\Delta$. The
parameters are $\Delta_2/\Delta_1=0.9$ and $J/\Delta_1=0.01$. The
dashed line represents a possible threshold fidelity for
identifying the minimum gate time. Here this threshold is set at
0.98.}
\label{Fig:SpeedLimitsCNOT}
\end{figure}

\textbf{Numerical calculations}: Here we use numerical
calculations to find the speed limits for a number of standard
two-qubit gates. The method is based on optimal control theory for
finding driving pulses that maximize the gate fidelity
\cite{Khaneja}. The fidelity is essentially a measure of the
overlap between the numerically calculated gate and the desired
target gate. Here we use the definition
\begin{equation}
{\rm Fidelity} = \left| \frac{ {\rm Tr} \left\{ U_{\rm
Target}^{\dagger} U_{\rm Numerical} \right\}}{2^n} \right|^2,
\end{equation}
where $U_{\rm Target}$ and $U_{\rm Numerical}$ are, respectively,
the target gate and any given unitary operation (which is obtained
from solving the Schr\"odinger equation with a given driving
signal), and $n$ is the number of qubits (here $n=2$). When the
gate time (which is a variable parameter in the calculations) is
set to a small value, the maximum achievable fidelity is
substantially smaller than unity. As the allowed time is
increased, the maximum achievable fidelity increases, until at a
certain value of the allowed time the fidelity reaches the value
one and remains at that value for larger times
\cite{SchulteHerbrueggen,Caneva}. In other words, when plotted as
a function of the allowed time, the fidelity exhibits non-analytic
behaviour as it suddenly hits the value one and remains there. The
time at which the fidelity reaches unity defines the minimum gate
time, which can alternatively be expressed as the speed limit. A
commonly used procedure for deducing speed limits is to set a
threshold value for the fidelity (say 99\%) and numerically
identify the minimum time required in order to attain this
fidelity. This procedure is illustrated in
Fig.~\ref{Fig:SpeedLimitsCNOT}.

\begin{figure}[h]
\includegraphics[width=8.0cm]{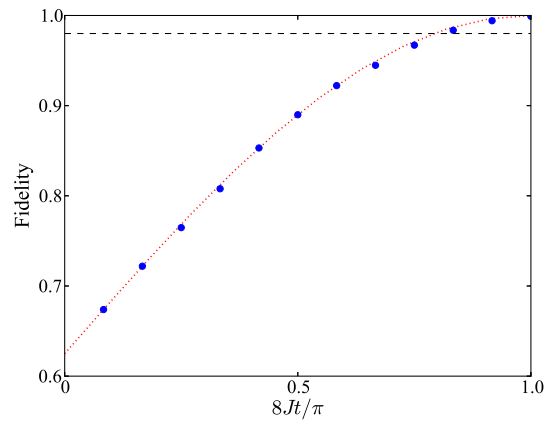}
\caption{(Color online) The fidelity of the $\sqrt{\rm SWAP}$ gate
for the pulse obtained using optimal control theory as a function
of the allowed time $t$ (plotted in the combination $8Jt/\pi$) in
the case of Heisenberg interactions and fixed values of $\Delta$.
The parameters are $\Delta_2/\Delta_1=0.9$ and $J/\Delta_1=0.01$.
The dashed line represents, as an example, a threshold fidelity of
0.98. The red dotted line is the function
$(5/8)+(3/8)\times\sin(4Jt)$. Note that the value $5/8$ is the
fidelity of the unit matrix with the $\sqrt{\rm SWAP}$ gate.}
\label{Fig:SpeedLimitsSQRTSWAP}
\end{figure}

Here we use an alternative procedure that avoids one of the
drawbacks of the above procedure, namely the slow convergence of
the pulse-optimization algorithm when the fidelity approaches its
asymptotic value (when plotted as a function of iteration number).
We calculate the maximum achievable fidelity for times varying
between zero and an estimated value for the minimum gate time. The
results of such a calculation are plotted in
Fig.~\ref{Fig:SpeedLimitsSQRTSWAP} for the case of the $\sqrt{\rm
SWAP}$ gate with Heisenberg interactions. We can see that the
results fit very well with a function that gives the gate time
$\pi/(8J)$. The sine function used in the figure was used as a
`trial' fitting function; it was inspired by the behaviour of
single-qubit gates but turned out to produce an excellent fit in
this case as well. We note here that this alternative method
(i.e.~the method where one looks away from the threshold region in
order to identify the minimum gate time) has not been used in the
literature in the context of optimal-control theory. More
complicated fitting functions might be required when applying this
method to systems with more than two qubits. However, it would be
interesting to explore in the future the usefulness of this
approach to similar problems. We shall use this method when
identifying the minimum gate time in the calculations below.

It is worth mentioning that the numerical method used here is
guaranteed to give the optimal pulses, and therefore the correct
speed limit, if given sufficient computation time \cite{Brif}.
Small-scale calculations could produce results that overestimate
the speed limit, because the algorithm might not converge to the
optimal pulses with the given calculation parameters. The
availability of known speed limits in the literature (e.g.~in
Ref.~\cite{Vidal}) for some physical setups and gates allows us to
characterize the performance of our calculations with a given set
of parameters. We find that in most cases we obtain the correct
speed limits to within a few percent with relatively small-scale
calculations.

The results for the deduced gate times are summarized in Table
\ref{Table:TwoQubitGates}. The numerical calculations are
performed with 500 time steps and $10^4$ iterations. With these
parameters a single data point takes a calculation time on the
order of one hour. We run each calculation a few times with a
variety of initial pulses in order to minimize the chances of slow
convergence towards the optimal pulse, which is a possibility
given the fact that we do not know the structure of the fidelity
landscape. For fixed-$\Delta$ calculations, we set
$\Delta_2/\Delta_1=0.9$ and $J/\Delta_1=0.01$. The speed limits
for the case of tunable $\Delta_i$ are known, and using them we
can see that our relatively small-scale calculations produce a
very good approximation to the speed limit in most cases. This
observation gives us confidence in the convergence behaviour of
the calculations.

The main observation that we make from the results in Table
\ref{Table:TwoQubitGates} is that all the results remained
unchanged whether the parameters $\Delta_i$ were taken to be fixed
or tunable, a result that is not {\it a prioi} obvious. In some
cases, the calculations with tunable $\Delta$ produced higher
estimates for the minimum gate time than the calculations with
fixed $\Delta$, even though the tunable-$\Delta$ case has more
tunable parameters and must therefore result in gates that are at
least as fast as those obtained in the fixed-$\Delta$ case. It
seems, however, that the presence of additional tunable parameters
can have the effect of slowing down the convergence of the
optimization algorithm, which we see for some of the cases
considered in Table \ref{Table:TwoQubitGates}. On the other hand,
one case where the fixed-$\Delta$ result is substantially higher
than the tunable-$\Delta$ result is that of the iSWAP gate with
Ising interactions. The speed limit that we find numerically for
the fixed-$\Delta$ case is about 25\% higher than that for the
tunable $\Delta$ case. However, this difference is again caused by
numerical inaccuracies. In the case where
$J\ll|\Delta_1-\Delta_2|\ll\Delta_1$ it is known that one can
approach a gate time of $\pi/(2J)$, and this fast gate is achieved
by strongly driving the two qubits such that they are effectively
tuned into resonance with each other \cite{Ashhab}. Since the
fixed-$\Delta$ speed limit cannot be higher than the
tunable-$\Delta$ speed limit, one can conclude that the speed
limit is $\pi/(2J)$ in both cases. The fact that a
large-amplitude, high-frequency driving pulse is needed in order
to achieve the speed limit (assuming that this is the only way to
achieve the speed limit) might partly explain why the algorithm is
not converging to the optimal pulses. Another partial explanation
of the relatively large gate time obtained in this case could be
the fact that $J/\Delta_1=0.01$, which could increase the minimum
gate time by a few percent.

It is also worth noting here that even though the Heisenberg
interaction Hamiltonian has more terms than the Ising interaction
Hamiltonian, which generally results in larger frequency shifts
and gaps in spectroscopy measurements, it is not the case that
Heisenberg interactions will always result in faster two-qubit
gates. One case that might be particularly surprising at first
sight is that of the iSWAP gate. The term proportional to $(
\hat{\sigma}_+^{(1)} \otimes \hat{\sigma}_-^{(2)} +
\hat{\sigma}_-^{(1)} \otimes \hat{\sigma}_+^{(2)} )$ is larger in
the case of Heisenberg interactions. However, the presence of the
$\hat{\sigma}_z^{(1)} \otimes \hat{\sigma}_z^{(2)}$ term in some
sense has the effect of slowing down the iSWAP-gate dynamics, such
that Heisenberg interactions and Ising interactions give the same
minimum gate times.

\begin{table*}[ht]
\caption{Minimum times required for various two-qubit gates and
physical realizations. The analytical expressions given below are
extracted from the numerical calculations: for each calculation we
identify the value of the time that gives a good fit to a figure
similar to Fig.~\ref{Fig:SpeedLimitsSQRTSWAP} (the sine function
used for the fidelity dependence always provided a good fit to the
data). In the fixed-$\Delta$ calculations, we set
$\Delta_2/\Delta_1=0.9$ and $J/\Delta_1=0.01$.}
\centering
\begin{tabular}{ccccccccc}
Gate & & CNOT & & CZ & & iSWAP & & $\sqrt{\rm SWAP}$ \\
\\
Matrix & & $\left( \begin{array}{cccc} 1 & 0 & 0 & 0 \\ 0 & 1 & 0
& 0 \\ 0 & 0 & 0 & 1 \\ 0 & 0 & 1 & 0 \end{array} \right)$ & &
$\left(
\begin{array}{cccc} 1 & 0 & 0 & 0 \\ 0 & 1 & 0 & 0 \\ 0 & 0 & 1 & 0 \\
0 & 0 & 0 & -1 \end{array} \right)$ & & $\left(
\begin{array}{cccc} 1 & 0 & 0 & 0 \\ 0 & 0 & i & 0 \\ 0 & i & 0 &
0 \\ 0 & 0 & 0 & 1
\end{array} \right)$ & & $\left( \begin{array}{cccc} 1 & 0 & 0 & 0 \\
0 & \frac{1+i}{2} & \frac{1-i}{2} & 0 \\ 0 & \frac{1-i}{2} & \frac{1+i}{2} & 0 \\
0 & 0 & 0 & 1 \end{array} \right)$ \\
\\
Ising, fixed $\Delta$ & & $\displaystyle\frac{\pi}{4J} \times$
1.12 & & $\displaystyle\frac{\pi}{4J} \times$ 1.14 & &
$\displaystyle\frac{\pi}{2J} \times$ 1.25 &
& $\displaystyle\frac{3\pi}{8J} \times$ 1.00 \\
\\
Ising, tunable $\Delta$ & & $\displaystyle\frac{\pi}{4J} \times$
1.17 & & $\displaystyle\frac{\pi}{4J} \times$ 1.26 & &
$\displaystyle\frac{\pi}{2J} \times$ 1.04 &
& $\displaystyle\frac{3\pi}{8J} \times$ 1.03 \\
\\
Heisenberg, fixed $\Delta$ & & $\displaystyle\frac{\pi}{4J}
\times$ 1.01 & & $\displaystyle\frac{\pi}{4J} \times$ 1.00 & &
$\displaystyle\frac{\pi}{2J} \times$ 1.04 &
& $\displaystyle\frac{\pi}{8J} \times$ 1.01 \\
\\
Heisenberg, tunable $\Delta$ & & $\displaystyle\frac{\pi}{4J}
\times$ 1.10 & & $\displaystyle\frac{\pi}{4J} \times$ 1.00 & &
$\displaystyle\frac{\pi}{2J} \times$ 1.02 & &
$\displaystyle\frac{\pi}{8J} \times$ 1.00
\end{tabular}
\label{Table:TwoQubitGates}
\end{table*}

\section{Three-qubit gates}
\label{Sec:ThreeQubitGates}

The most well-known three-qubit gate is the Toffoli gate. This
gate is also known as the controlled-controlled-NOT gate, because
it applies the NOT operation to the target qubit when both control
qubits are in the state $\ket{1}$. It has been known for some time
that the Toffoli gate can be constructed from six CNOT gates, in
addition to a number of single-qubit gates. Recently it has been
shown that this number (i.e.~six) is the minimum number of CNOT
gates required in order to construct the Toffoli gate
\cite{Shende}. Different constructions based on general
conditional gates have also been proposed, reducing the
Toffoli-gate time from 6 to 3.5 times that CNOT gate time
\cite{Barenco}. An optimal-control-theory-related calculation
based on certain forms of pulses has found a gate time of about
2.2 times the CNOT gate time for qubits coupled in a triangle
geometry \cite{Yuan}.

We have performed numerical calculations in order to determine the
minimum time required in order to obtain the Toffoli gate for a
Hamiltonian with pairwise interaction terms given by
Eq.~(\ref{Eq:IsingCouplingHamiltonian}) or
Eq.~(\ref{Eq:HeisenbergCouplingHamiltonian}). We should stress
here that these calculations are independent of the results
mentioned above. The reason is that the gate-counting calculations
assume that the different gates are applied as separate,
well-defined units, and in some calculations it is assumed that
the two-qubit gates used in the construction are taken from a
specific class of gates (e.g.~conditional gates). One can
therefore expect that it should be possible to obtain a shorter
minimum time for the Toffoli gate by considering essentially all
the possible driving pulses.

The results are summarized in Table \ref{Table:ToffoliGate}. There
we show results where all coupling strengths are equal (allowing
small differences between the different coupling strengths does
not change the main results). We only present results for
fixed-$\Delta$ calculations, because the variable-$\Delta$
calculations did not produce any useful results within any
reasonable calculation time. In addition to the fact that our
procedure typically overestimates the minimum gate time, we would
like to point out that the results of the calculations also showed
stronger dependence on the guess pulses than in the case of
two-qubit gates (the data also did not result in smooth fitting
functions as shown in Fig.~\ref{Fig:SpeedLimitsSQRTSWAP}). We
therefore cannot exclude the possibility that some of the
expressions in Table \ref{Table:ToffoliGate} are substantially
larger than the true minimum gate time. However, for the triangle
geometry, we find Toffoli gate times that are smaller than twice
the minimum time required for a CNOT gate with the same coupling
strength. In this case, we can have a good level of confidence
that the results are at least close to the true minimum gate time,
since it seems implausible that the Toffoli gate could be faster
than the CNOT gate for the same value of the coupling strengths.
From the results shown in the table, we can also conclude that
having the qubits connected in a triangle geometry would be
desirable for purposes of implementing fast three-qubit gates. We
should point out here that the results presented here give faster
gates than any results for the maximum speed of Toffoli gates in
the literature \cite{Barenco,Yuan}. In particular, one can compare
the factor 1.9 in Table \ref{Table:ToffoliGate} with the
corresponding factor of 2.2 in Ref.~\cite{Yuan}: an improvement of
about 15\%.

\begin{table}[ht]
\caption{Minimum times required for the three-qubit Toffoli gate
with various physical realizations. The results are given in terms
of the CNOT gate time, i.e.~$\pi/(4J)$. The numerical calculations
are performed with $\sim 10^4$ time steps and $\sim 10^4$
iterations (a single data point now takes a calculation time on
the order of one day). The parameters $\Delta_i$ are fixed in the
calculations, and we set $\Delta_2/\Delta_1=0.9$,
$\Delta_3/\Delta_1=0.82$ and $J/\Delta_1=0.01$ for all of the
coupling terms.}
\centering
\begin{tabular}{ccccccc}
Geometry & & chain & & chain & & triangle \\
Target qubit & & center qubit & & side qubit & & any qubit \\
Ising & & 3.8 & & 3.8 & & 1.9 \\
Heisenberg & & 2.8 & & 2.6 & & 1.4
\end{tabular}
\label{Table:ToffoliGate}
\end{table}

\section{Quantum state transfer in a chain of qubits}
\label{Sec:StateTransfer}

Another operation in multi-qubit systems that has received
considerable attention in recent years is that of quantum state
transfer across a chain of qubits \cite{SpinChainPapers}. In this
section we provide analytical expressions for the speed limit of
state transfer across a long chain.

The Hamiltonian for a qubit chain with nearest neighbour
interactions is given by
\begin{equation}
\hat{H} = \sum_{i=1}^{N} \left( - \frac{\Delta_i}{2}
\hat{\sigma}_x^{(i)} - \frac{\epsilon_i}{2} \hat{\sigma}_z^{(i)}
\right) + \sum_{i=1}^{N-1} J_i \hat{\sigma}_z^{(i)} \otimes
\hat{\sigma}_z^{(i+1)}
\label{Eq:IsingCouplingHamiltonianChain}
\end{equation}
for the case of Ising interactions and
\begin{eqnarray}
\hat{H} & = & \sum_{i=1}^{N} \left( - \frac{\Delta_i}{2}
\hat{\sigma}_x^{(i)} - \frac{\epsilon_i}{2} \hat{\sigma}_z^{(i)}
\right) + \sum_{i=1}^{N-1} J_i \times \nonumber \\ & & \left(
\hat{\sigma}_x^{(i)} \otimes \hat{\sigma}_x^{(i+1)} +
\hat{\sigma}_y^{(i)} \otimes \hat{\sigma}_y^{(i+1)} +
\hat{\sigma}_z^{(i)} \otimes \hat{\sigma}_z^{(i+1)} \right).
\label{Eq:HeisenbergCouplingHamiltonianChain}
\end{eqnarray}
for Heisenberg interactions. The case with fixed and generally
disordered values of $\Delta_i$ and/or $J_i$ leads to serious
complications (such as Anderson localization), and we therefore do
not consider this case. For the case of fixed, uniform values of
$\Delta_i$ and $J_i$ or the case of tunable $\Delta_i$ and uniform
$J_i$, we can use arguments from band theory to find the speed
limit for state transfer.

The state-transfer process can be thought of as the process of
wave propagation through the chain, as discussed in
Ref.~\cite{Osborne}. The speed limit is then determined by the
maximum group velocity of a wave packet traveling through the
chain. We also observe here that wave propagation cannot be sped
up through the application of nonuniform external fields. We can
therefore calculate the maximum wave speed assuming a uniform
external field (i.e.~time- and position-independent values of
$\Delta_i$ and $\epsilon_i$). For the case of Ising interactions,
we start by noting that if we take $\Delta=0$, then the operator
$\hat{\sigma}_z^{(i)}$ commutes with the Hamiltonian, prohibiting
``propagation''. We therefore conclude that the condition
$\Delta\gg J$ is optimal for maximizing the propagation speed.
With this assumption, the wave function that describes a wave with
momentum $k$ (with $-\pi<k<\pi$) is given by
\begin{equation}
\ket{k} = \sum_{j=1}^{N} e^{ikj} \hat{\sigma}_z^{(j)}
\ket{\sigma_x^{(1)}=1, \cdots, \sigma_x^{(N)}=1}.
\label{Eq:SpinWaveFunction}
\end{equation}
The energy spectrum of these waves is given by $E_k = 2 J \cos k$.
The group velocity of the wave is given by
\begin{eqnarray}
v_k = \frac{dE}{dk} = - 2 J \sin k,
\label{Eq:IsingCouplingGroupVelocity}
\end{eqnarray}
which has a maximum at $k=\pi/2$, and the maximum is given by
$2J$. The minimum time for the wave to traverse a chain of length
$N$ is attained when the wave is initialized and set to travel at
this maximum wave speed. Ignoring chain-length-independent
contributions at the beginning and end of the transfer process,
the minimum transfer time is therefore given by
\begin{equation}
T_{\rm min} = \frac{N}{2J}.
\end{equation}
It is worth noting that $T_{\rm min}$ is faster by a factor of
$\pi$ than the sequential application of iSWAP operations across
the chain. We also note that the expression for the maximum wave
speed given above can be found in the literature, e.g.~in
Ref.~\cite{Happola}.

For the case of Heisenberg interactions, the speed of wave
propagation is independent of the external fields. Taking
$\Delta\gg J$, $\epsilon=0$, and assuming only one excitation in
the system, the problem reduces to that with Ising interactions
but with twice the coupling strength: the $\hat{\sigma}_x^{(i)}
\otimes \hat{\sigma}_x^{(j)}$ term has no effect on the dynamics,
and
\begin{equation}
\hat{\sigma}_y^{(i)} \otimes \hat{\sigma}_y^{(j)} +
\hat{\sigma}_z^{(i)} \otimes \hat{\sigma}_z^{(j)}= 2 \left(
\hat{\sigma}_+^{(i)} \otimes \hat{\sigma}_-^{(j)} +
\hat{\sigma}_-^{(i)} \otimes \hat{\sigma}_+^{(j)} \right).
\end{equation}
The minimum transfer time is therefore given by
\begin{equation}
T_{\rm min} = \frac{N}{4J}.
\end{equation}
This expression is essentially the same as the one given in
Ref.~\cite{Osborne}. Using numerical calculations
Ref.~\cite{Caneva} found a time that is a few percent higher (note
that there is a factor-of-two difference between
Eq.~(\ref{Eq:HeisenbergCouplingHamiltonianChain}) and the
Hamiltonian used in Ref.~\cite{Caneva}).

Even though the maximum wave speed sets an upper limit to the
speed of quantum-state transfer, one still needs to make sure that
the quantum state is transferred without distortion. The authors
of Ref.~\cite{Osborne} assumed that one has access to a limited
number of qubits in the chain and analyzed the dispersion of the
propagating wave. They found that if one has access to a number
that is at least on the order of $\sqrt{N}$ at both ends of the
chain, then state transfer can occur with high fidelity. The
authors of Ref.~\cite{Caneva} assumed access to all of the qubits
and, using numerical calculations, demonstrated that by using a
`carrier' external field (e.g. a harmonic-oscillator potential
that moves along the chain and carries the quantum state as it
moves along) the dispersion of the wave can also be prevented, and
high-fidelity state transfer is possible.

\section{Conclusion}
\label{Sec:Conclusion}

We have derived a number of speed limits, or lower bounds on the
required time, for various quantum operations in various setups
that correspond to a variety of experimental conditions.
Single-qubit operation speeds are limited by the smaller of the
Pauli-matrix coefficients in the Hamiltonian. We have used optimal
control theory to obtain speed limits for a few well-known
two-qubit gates and the three-qubit Toffoli gate. As expected,
two-qubit gate speeds are limited by the inter-qubit coupling
strength. The Toffoli gate requires approximately three times the
minimum time required for a CNOT gate in a chain geometry and less
than twice the minimum time required for a CNOT gate in a triangle
geometry. Finally, we have used arguments from condensed-matter
physics to derive the speed limit for quantum state transfer in a
qubit chain. The expressions that we find agree with analytical
results known in the literature and also with recent numerical
results.

\section*{Acknowledgments}

We would like to thank M. R. Geller, J. R. Johansson, A. Lupascu,
R. Wu and F. Xue for useful discussions. This work was supported
in part by LPS, NSA, ARO, NSF Grant No. 0726909, JSPS-RFBR
Contract No. 09-02-92114, Grant-in-Aid for Scientific Research
(S), MEXT Kakenhi on Quantum Cybernetics, and the JSPS via its
FIRST program.

\end{document}